# Entanglement generation and quantum information transfer between spatially-separated qubits in different cavities

Chui-Ping Yang[1,2,3], Qi-Ping Su[3], and Franco Nori[1,2]

[1]*CEMS, RIKEN, Saitama, 351-0198, Japan*

[2]*Physics Department, The University of Michigan, Ann Arbor, MI 48109-1040, USA and*

[3]*Department of Physics, Hangzhou Normal University, Hangzhou, Zhejiang 310036, China*

(Dated: July 9, 2013)

The generation and control of quantum states of spatially-separated qubits distributed in different cavities constitute fundamental tasks in cavity quantum electrodynamics. An interesting question in this context is how to prepare entanglement and realize quantum information transfer between qubits located at different cavities, which are important in large-scale quantum information processing. In this paper, we consider a physical system consisting of two cavities and three qubits. Two of the qubits are placed in two different cavities while the remaining one acts as a coupler, which is used to connect the two cavities. We propose an approach for generating quantum entanglement and implementing quantum information transfer between the two spatially-separated intercavity qubits. The quantum operations involved in this proposal are performed by a virtual photon process, and thus the cavity decay is greatly suppressed during the operations. In addition, to complete the present tasks, only one coupler qubit and one operation step are needed. Moreover, there is no need of applying classical pulses, so that the engineering complexity is much reduced and the operation procedure is greatly simplified. Finally, our numerical results illustrate that high-fidelity implementation of this proposal using superconducting phase qubits and one-dimenstion transmision line resonators is feasible for current circuit QED implementations. This proposal can also be applied to other types of superconducting qubits, including flux and charge qubits.

## I. INTRODUCTION

There exist several physical systems in which a quantum bus could be realized. One example is trapped ions [1,2], in which various quantum operations and algorithms have been performed by employing the quantized motion of the ions (phonons) as the bus. Photons are highly coherent and can mediate interactions between distant objects, and thus are another natural candidate as a carrier of quantum information [3,4]. A photon bus can be created by using an atom coupled to a single-cavity mode via cavity quantum electrodynamics (QED). In the strong coupling limit [5], the interaction between the atom and the cavity mode is coherent, allowing the transfer of quantum information between the atom and the photon. The experimental demonstration of entanglement between atoms has been reported with Rydberg-atom cavity QED [6-8]. In addition, using photons transmitted via a transmission line (e.g., an optical fiber), the transfer of quantum information or quantum states from one atom to another distant atom was previously considered [9] and has been extensively studied [10]. Moreover, a quantum network based on single atoms placed in optical cavities, which are coupled by optical fibers, has been proposed [11], and the transfer of an atomic quantum state and the creation of entanglement between two nodes in such a network has been experimentally demonstrated [11]. As is well known, entanglement and quantum information transfer have played a central role in the field of quantum information due to their potential applications in quantum cryptography, quantum communication, quantum computing, and so on.

Superconducting devices [12-14] play important roles in quantum information processing (QIP). Circuit QED is a realization of the physics of cavity QED with superconducting qubits coupled to a microwave cavity on a chip, and has been considered as one of the most promising candidates for QIP [12,13]. Previous circuit QED experiments have achieved the strong-coupling limit with a superconducting qubit coupled to a cavity [15,16]. Based on circuit QED, many theoretical works have studied the preparation of Fock states, coherent states, squeezed states, Schrödinger cat states, and an arbitrary superposition of Fock states of a single superconducting cavity [17-20]. Also, the experimental creation of a Fock state and a superposition of Fock states of a single superconducting cavity using a superconducting qubit has been reported [21,22]. Moreover, a large number of theoretical proposals have been presented for realizing quantum information transfer, logical gates, and entanglement with two or more superconducting qubits embedded in a cavity or coupled by a resonator [23-32]. Hereafter, we use the term cavity and resonator interchangeably. In addition, quantum information transfer, two-qubit gates, three-qubit gates and three-qubit entanglement have been experimentally demonstrated with superconducting qubits in a single cavity [33-37]. However, large-scale QIP will need many qubits and placing all of them in a single cavity could cause many fundamental and practical problems, e.g., increasing the cavity decay rate and decreasing the qubit-cavity coupling strength.

Considerable experimental and theoretical work has been devoted recently to the investigation of QIP in a system consisting of two or more than two cavities, each hosting (and coupled) to multiple qubits. In this kind of architecture, quantum operations would be performed not only on qubits in the same cavity, but also on qubits or photons in different cavities. Within circuit QED, several theoretical proposals for generation of entangled photon Fock states of two resonators have been presented [38,39]. Reference [40] proposed a theoretical scheme for creating NOON states of two resonators, which has been implemented in experiments [41]. Moreover, schemes for preparation of entangled photon Fock states or entangled coherent states of more than two cavities have been presented recently [42-44].

In the following, we consider a physical system in which two cavities are interconnected to a superconducting coupler qubit and each cavity hosts a superconducting qubit. Our goal is to propose an approach for generating quantum entanglement and quantum information transfer between the two spatially-separated intercavity qubits. As shown below, the quantum operations involved in this proposal are carried out by a virtual photon process (i.e., photons of the cavity modes are not populated or excited). Hence, the cavity decay is greatly suppressed during the operations. In addition, the present proposal has several distinguishing features: only one superconducting coupler qubit and one operation step are needed, and no classical microwave pulse is used during the operation, so that the circuit complexity is much reduced and the operation procedure is greatly simplified.

The method presented here is quite general, and can be applied to accomplish the same task with the coupler qubit replaced by a different type of qubit such as a quantum dot, or with the two intercavity qubits replaced by other two qubits such as two atoms, two quantum dots, two NV centers and so on.

This paper is organized as follows. In Sec. 2, we show how to generate quantum entanglement and perform quantum information transfer between two superconducting qubits located at two different cavities, and then give a brief discussion on the experimental issues. In Sec. 3, we present a brief discussion of the fidelity and possible experimental implementation with superconducting phase qutrits as an example. A concluding summary is enclosed in Sec. 4.

## II. ENTANGLEMENT AND INFORMATION TRANSFER

Consider two cavities 1 and 2 coupled by a two-level superconducting qubit $A$, as illustrated in Fig. 1(a). Cavity 1 hosts a two-level superconducting qubit 1, shown as a black dot, and cavity 2 hosts another two-level superconducting qubit 2. Each qubit here has two levels, $|0\rangle$ and $|1\rangle$. We here assume that the coupling constant of qubit 1 with cavity 1 is $g_1$ and the coupling constant of qubit 2 with cavity 2 is $g_2$. The coupler qubit $A$ in Fig. 1 can interact with both cavities 1 and 2 simultaneously, through the qubit-cavity capacitors $C_1$ and $C_2$. We denote $g_{A1}$ as the coupling constant of qubit $A$ with cavity 1 and $g_{A2}$ as the coupling constant of qubit $A$ with cavity 2. In the interaction picture, we have

$$H_I = \sum_{j=1}^{2} g_j \left( e^{i\delta_j t} a_j \sigma_j^+ + h.c. \right) + \sum_{j=1}^{2} g_{Aj} \left( e^{i\delta_{Aj} t} a_j \sigma_A^+ + h.c. \right), \qquad (1)$$

where $\sigma_j^+ = |1\rangle_j \langle 0|$ ($\sigma_A^+ = |1\rangle_j \langle 0|$) is the raising operator for qubit $j$ (qubit $A$), $\delta_j = \omega_{10j} - \omega_{cj}$ is the detuning of the transition frequency $\omega_{10j}$ of qubit $j$ from the frequency $\omega_{cj}$ of cavity $j$, $\delta_{Aj} = \omega_{10A} - \omega_{cj}$ is the detuning of the transition frequency $\omega_{10A}$ of qubit $A$ from frequency $\omega_{cj}$ of cavity $j$ [Fig. 1(b,c,d)], and $a_j$ is the annihilation operator for the mode of cavity $j$ ($j = 1, 2$).

Assuming $\delta_j \gg g_j$, $\delta_{Aj} \gg g_{Aj}$ and under the condition of

$$|\delta_{A2} - \delta_{A1}| \gg \frac{g_{A1} g_{A2}}{2} \left( 1/\delta_{A1} + 1/\delta_{A2} \right), \qquad (2)$$

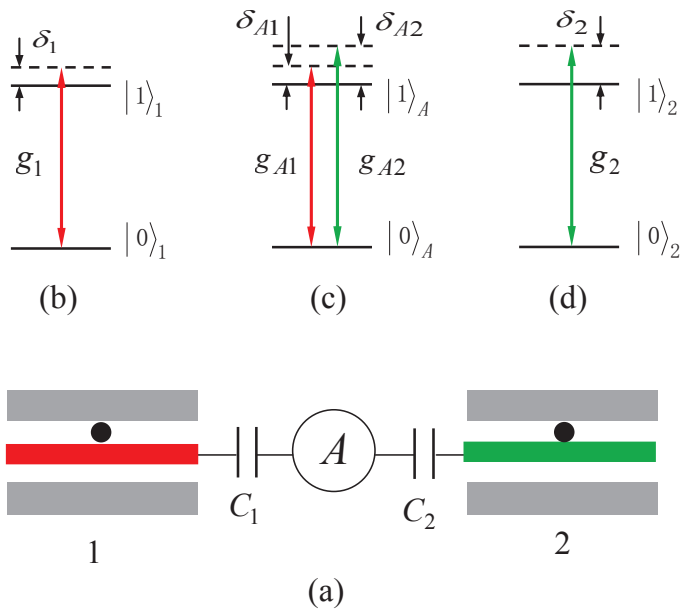

FIG. 1: (Color online) (a) Setup for two cavities 1 and 2 coupled by a superconducting qubit $A$. Each cavity here is a one-dimensional coplanar waveguide transmission line resonator. The circle $A$ represents a superconducting qubit, which is capacitively coupled to cavity $j$ via a capacitance $C_j$ $(j=1,2)$. The two dark dots indicate the two superconducting qubits 1 and 2 embedded in the two cavities, respectively. (b) Illustration of qubit 1 dispersively interacting with cavity 1. (c) Illustration of the coupler qubit $A$ dispersively interacting with both cavities 1 and 2. (d) Illustration of qubit 2 dispersively coupled to cavity 2.

we can obtain [7,45]

$$\begin{aligned} H_{\text{eff}} &= -\sum_{j=1}^{2} \frac{g_j^2}{\delta_j} \left( |0\rangle_j \langle 0| a_j^+ a_j - |1\rangle_j \langle 1| a_j a_j^+ \right) \\ &\quad - \sum_{j=1}^{2} \frac{g_{Aj}^2}{\delta_{Aj}} \left( |0\rangle_A \langle 0| a_j^+ a_j - |1\rangle_A \langle 1| \right) a_j a_j^+ \\ &\quad + \sum_{j=1}^{2} \lambda_j \left( e^{i(\delta_j - \delta_{Aj})t} \sigma_j^+ \sigma_A + h.c. \right) \end{aligned} \quad (3)$$

where $\lambda_j = \frac{g_j g_{Aj}}{2} (1/\delta_j + 1/\delta_{Aj})$.

Assume that the two cavities are initially in the vacuum state, and set

$$\delta_1 = \delta_{A1}, \ \delta_2 = \delta_{A2}. \quad (4)$$

Then the Hamiltonian (3) reduces to

$$H_{\text{eff}} = H_0 + H_{\text{int}}, \quad (5)$$

with

$$\begin{aligned} H_0 &= \sum_{j=1}^{2} \frac{g_j^2}{\delta_j} |1\rangle_j \langle 1| + \sum_{j=1}^{2} \frac{g_{Aj}^2}{\delta_{Aj}} |1\rangle_A \langle 1|, \\ H_{\text{int}} &= \sum_{j=1}^{2} \lambda_j \left( \sigma_j^+ \sigma_A + \sigma_j \sigma_A^+ \right). \end{aligned} \quad (6)$$

In a new interaction picture under the Hamiltonian $H_0$, and under the following condition

$$\frac{g_1^2}{\delta_1} = \frac{g_2^2}{\delta_2} = \sum_{j=1}^{2} \frac{g_{Aj}^2}{\delta_{Aj}}, \quad (7)$$

we can obtain

$$\widetilde{H}_{\text{int}} = e^{iH_0 t} H_{\text{int}} e^{-iH_0 t} = H_{\text{int}}. \quad (8)$$

Based on this Hamiltonian and after returning to the original interaction picture by performing a unitary transformation $e^{-iH_0 t}$, one can easily find the following state evolution

$$
\begin{aligned}
|000\rangle &\rightarrow |000\rangle \\
|100\rangle &\rightarrow N \left[ e^{-i\varphi_1 t} \left( 1 + \frac{\lambda_1^2}{\lambda_2^2} \cos \Lambda t \right) |100\rangle + e^{-i\varphi_2 t} \frac{\lambda_1}{\lambda_2} (\cos \Lambda t - 1) |010\rangle \right] \\
&\quad - i\sqrt{N} e^{-i\varphi_A t} \frac{\lambda_1}{\lambda_2} \sin \Lambda t |001\rangle, \\
|110\rangle &\rightarrow e^{-i(\varphi_1 + \varphi_2) t} \cos \Lambda t |110\rangle \\
&\quad - i\sqrt{N} \left[ e^{-i(\varphi_2 + \varphi_A) t} \frac{\lambda_1}{\lambda_2} \sin \Lambda t |011\rangle + e^{-i(\varphi_1 + \varphi_A) t} \sin \Lambda t |101\rangle \right],
\end{aligned}
\quad (9)
$$

where $\varphi_j = g_j^2/\delta_j$ $(j = 1, 2)$, $\varphi_A = \sum_{j=1}^{2} g_{Aj}^2/\delta_{Aj}$, $N = \lambda_2^2/(\lambda_1^2 + \lambda_2^2)$, and $\Lambda = \sqrt{\lambda_1^2 + \lambda_2^2}$. Here and below, $|ijk\rangle = |i\rangle_1 |j\rangle_2 |k\rangle_A$, with $i, j, k \in \{0, 1, 2\}$, and subscripts 1, 2, and $A$ indicating qubits 1, 2, and $A$ respectively.

The result (9) obtained here will be employed to create entanglement and to implement quantum information transfer between qubits 1 and 2, as shown below.

### A. Generation of entanglement

Initially, qubits 1, 2 and $A$ are in the state $|110\rangle$ and decouped from the two cavities by prior adjustment of each qubit's level spacings. Each cavity is initially in the vacuum state. For superconducting devices, the level spacings can be rapidly adjusted by varying external control parameters (e.g., magnetic flux applied to phase, transmon, or flux qutrits; see, e.g., [46,47]).

To generate the entanglement of qubits 1 and 2, we now adjust the qubit level spacings to achieve the state evolution described by Eq. (9). Under the condition (4) and the following condition

$$\frac{g_1^2}{\delta_1} = \frac{g_2^2}{\delta_2}, \quad g_{A1} = \frac{g_1}{\sqrt{2}}, \quad g_{A2} = \frac{g_2}{\sqrt{2}}, \quad (10)$$

one can verify $\lambda_1 = \lambda_2$ and $(\varphi_1 + \varphi_A) t_1 = (\varphi_2 + \varphi_A) t_1 = \pi$ for $t_1 = \pi/(2\Lambda)$. Using these results, one can see from Eq. (9) that after an interaction time $t_1 = \pi/(2\Lambda)$, the initial state $|110\rangle$ of the three qubits evolves into

$$|\phi\rangle = -i\frac{1}{\sqrt{2}} \left[|01\rangle + |10\rangle\right] |1\rangle, \quad (11)$$

which shows that the two qubits 1 and 2 are prepared in a maximally-entangled state, while the coupler qubit $A$ is left in the state $|1\rangle$. To freeze the prepared entangled state, the level spacings for each qubit need to be adjusted back to the original configuration, such that each qubit is decoupled from the two cavities.

### B. Transfer of quantum information

Suppose that qubit 1 is initially in an arbitrary state $\alpha |0\rangle + \beta |1\rangle$, qubits 2 and 3 are in the state $|00\rangle$, and each cavity is in the vacuum state. The three qubits are initially decoupled from each cavity by prior adjusting the qubit level spacings. Now, adjust the qubit level spacings to obtain the state evolution given in Eq. (9). It can be seen from Eq. (9) that after an interaction time $t_2 = \pi/\Lambda$, the initial state $(\alpha |0\rangle + \beta |1\rangle)|0\rangle|0\rangle$ of the qubit system changes to

$$\alpha |000\rangle + \beta e^{-i\varphi_1 t_2} N \left(1 - \frac{\lambda_1^2}{\lambda_2^2}\right) |100\rangle - \beta e^{-i\varphi_2 t_2} 2N \frac{\lambda_1}{\lambda_2} |010\rangle. \tag{12}$$

Under the conditions (4) and (10), we have $\lambda_1 = \lambda_2$ and $\varphi_2 t_2 = \pi$. Thus, the state (12) reduces to

$$|\varphi\rangle = |0\rangle (\alpha |0\rangle + \beta |1\rangle)|0\rangle. \tag{13}$$

Comparing the state (13) with the initial state of the qubit system, one can see that the following state tranformation is obtained, i.e.,

$$(\alpha |0\rangle + \beta |1\rangle)|0\rangle|0\rangle \to |0\rangle (\alpha |0\rangle + \beta |1\rangle)|0\rangle, \tag{14}$$

which demonstrates that the original quantum state (quantum informaton) of qubit 1 has been transferred onto qubit 2, while the coupler qubit $A$ remains in its original ground state $|0\rangle_A$. After completing the information transfer, one would need to adjust the qubit level spacings such that the qubits are decoupled from each cavity.

We should mention that adjusting the qubit level spacings is unnecessary. Alternatively, the coupling or decouping of the qubits with the cavities can be obtained by adjusting the frequency of each cavity. The rapid tuning of cavity frequencies has been demonstrated in superconducting microwave cavities (e.g., in less than a few nanoseconds for a superconducting transmission line resonator [48]).

For the method to work, the following requirements need to be satisfied:

(i) The conditions (4) and (10) need to be met. Here, note that the condition (7) is ensured by the condition (10). Also, $\delta_j$ and $\delta_{Aj}$ can be adjusted by varying the cavity frequency $\omega_{cj}$, the qubit transition frequency $\omega_{10j}$, or the coupler qubit transition frequency $\omega_{10A}$ ($j = 1, 2$). In addition, $g_{Aj}$ can be adjusted by changing the qubit-cavity coupler capacitancy $C_j$ (see Fig. 1). Hence, the conditions (4) and (10) can be readily satisfied.

(ii) The operation time required for the entanglement preparation or information transfer needs to be much shorter than the energy relaxation time $T_1$ and dephasing time $T_2$ of the level $|1\rangle$, such that the decoherence, caused by energy relaxation and dephasing of the qubits, is negligible during the operation.

(iii) For cavity $i$ ($i = 1, 2$), the lifetime of the cavity mode is given by $T_{\text{cav}}^i = (Q_i/2\pi\nu_{c,i})/\overline{n}_i$, where $Q_i$ and $\overline{n}_i$ are the (loaded) quality factor and the average photon number of cavity $i$, respectively. For the two cavities here, the lifetime of the cavity modes is given by

$$T_{\text{cav}} = \frac{1}{2} \min\{T_{\text{cav}}^1, T_{\text{cav}}^2\}, \tag{15}$$

which should be much longer than the operation time, such that the effect of cavity decay is negligible for the operation.

(iv) During the operation, there exists an intercavity cross coupling which is determined mostly by the coupling capacitances $C_1$ and $C_2$, and the qutrit's self capacitance $C_q$, because the field leakage through space is extremely low for high-$Q$ resonators as long as the inter-cavity distance is much greater than the transverse dimension of the cavities — a condition easily met in experiments for the two resonators. Furthermore, as our numerical simulations, shown by Figs. 3 and 4 below, the effects of the inter-cavity coupling can however be made negligible as long as $g_{12} \leq 0.2 g_{\max}$ with $g_{\max} = \max\{g_{A1}, g_{A2}\}$, where $g_{12}$ is the corresponding intercavity coupling constant between the two cavities.

## III. POSSIBLE EXPERIMENTAL IMPLEMENTATION

So far we have considered a general type of qubit. As an example of experimental implementation, let us now consider each qubit as a superconducting phase qubit. In reality, a third higher level $|2\rangle$ for each phase qubit here needs to be considered during the operations described above, since this level $|2\rangle$ may be occupied due to the $|1\rangle \leftrightarrow |2\rangle$ transition induced by the cavity mode(s), which will turn out to affect the operation fidelity. Hence, to quantify how well the proposed protocol works out, we will analyze the fidelity of the operation for both entanglement generation and information transfer, by considering a third higher level $|2\rangle$. Since three levels are now involved, we rename the three qubits 1, 2, and $A$ as qutrits 1, 2, and $A$, respectively.

When the intercavity crosstalk coupling and the unwanted $|1\rangle \leftrightarrow |2\rangle$ transition of each phase qutrit are considered, the Hamiltonian (1) is modified as follows

$$\widetilde{H}_I = H_I + H_I', \tag{16}$$

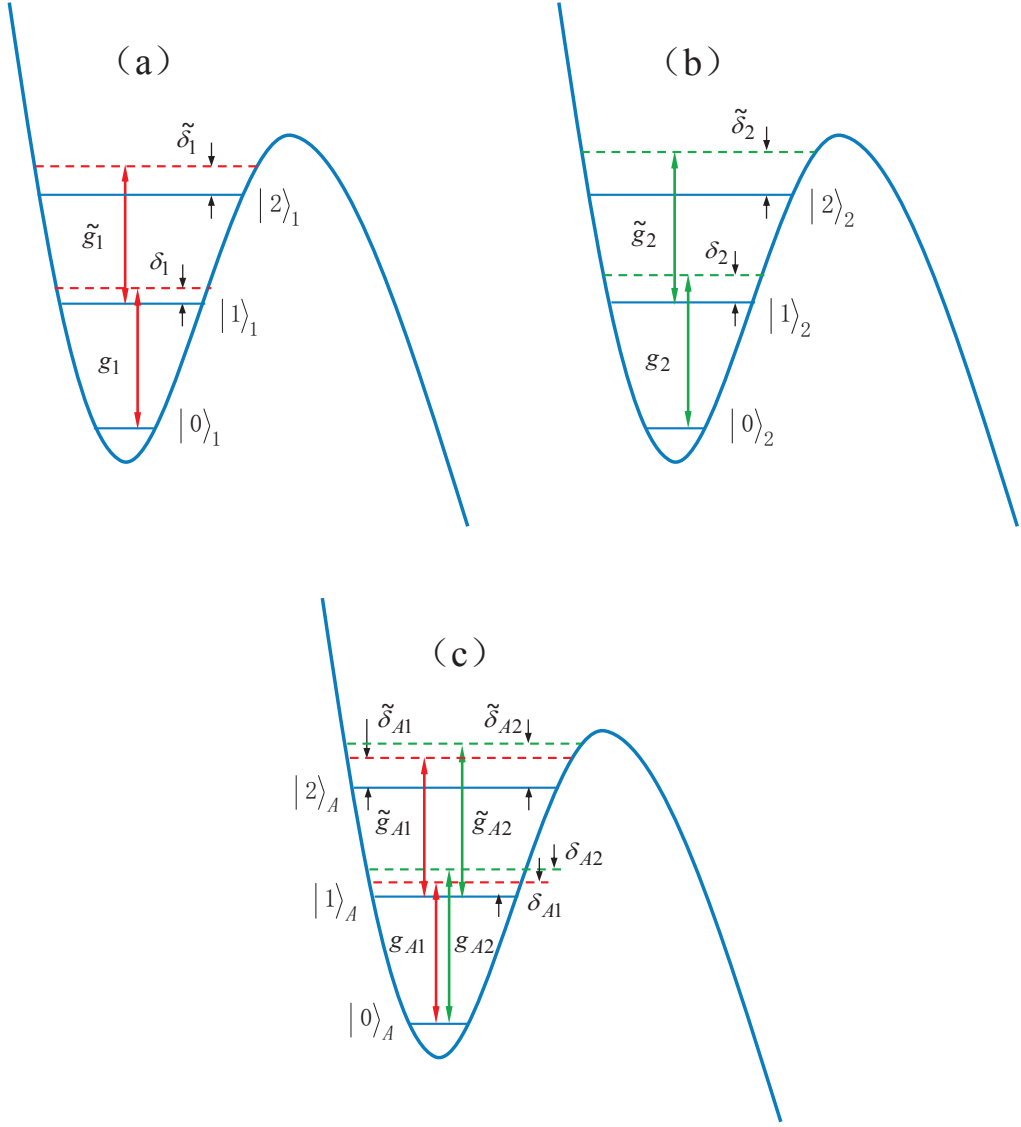

FIG. 2: (Color online) Illustration of qutrit-cavity interaction. (a) Cavity 1 is dispersively coupled to the $|0\rangle \leftrightarrow |1\rangle$ transition with coupling constant $g_1$ and detuning $\delta_1$, but far-off resonant (i.e., more detuned) with the $|1\rangle \leftrightarrow |2\rangle$ transition of qutrit 1 with coupling consant $\widetilde{g}_1$ and detuning $\widetilde{\delta}_1$. (b) Cavity 2 is dispersively coupled to the $|0\rangle \leftrightarrow |1\rangle$ transition with coupling constant $g_2$ and detuning $\delta_2$, but far-off resonant with the $|1\rangle \leftrightarrow |2\rangle$ transition of qutrit 2 with coupling consant $\widetilde{g}_2$ and detuning $\widetilde{\delta}_2$. (c) Cavity 1 (cavity 2) dispersively interacts with the $|0\rangle \leftrightarrow |1\rangle$ transition with coupling constant $g_{A1}$ ($g_{A2}$) and detuning $\delta_{A1}$ ($\delta_{A2}$), but is far-off resonant with the $|1\rangle \leftrightarrow |2\rangle$ transition of qutrit $A$ with coupling consant $\widetilde{g}_{A1}$ ($\widetilde{g}_{A2}$) and detuning $\widetilde{\delta}_{A1}$ ($\widetilde{\delta}_{A2}$). Here, $\delta_j = \omega_{10j} - \omega_{cj}, \widetilde{\delta}_j = \omega_{21j} - \omega_{cj}, \delta_{Aj} = \omega_{10A} - \omega_{cj}$, and $\widetilde{\delta}_{Aj} = \omega_{21A} - \omega_{cj}$ ($j = 1, 2$), where $\omega_{10j}$ ($\omega_{21j}$) is the $|0\rangle \leftrightarrow |1\rangle$ ($|1\rangle \leftrightarrow |2\rangle$) transition frequency of qutrit $j$, $\omega_{10A}$ ($\omega_{21A}$) is the $|0\rangle \leftrightarrow |1\rangle$ ($|1\rangle \leftrightarrow |2\rangle$) transition frequency of qutrit $A$, and $\omega_{cj}$ is the frequency of cavity $j$ ($j = 1, 2$).

where $H_I$ is the needed interaction Hamiltonian given in Eq. (1) above, while $H'_I$ is the unwanted interaction Hamiltonian, given by

$$H'_I = \sum_{j=1}^{2} \widetilde{g}_j \left( e^{i\widetilde{\delta}_j t} a_j \sigma^+_{21j} + h.c. \right) + \sum_{j=1}^{2} \widetilde{g}_{Aj} \left( e^{i\widetilde{\delta}_{Aj} t} a_j \sigma^+_{21A} + h.c. \right)$$
$$+ g_{12} \left( e^{i\Delta t} a_1 a_2^+ + h.c. \right), \tag{17}$$

where $\sigma^+_{21j} = |2\rangle_j \langle 1|$ and $\sigma^+_{21A} = |2\rangle_A \langle 1|$. The first term represents the unwanted off-resonant coupling between the

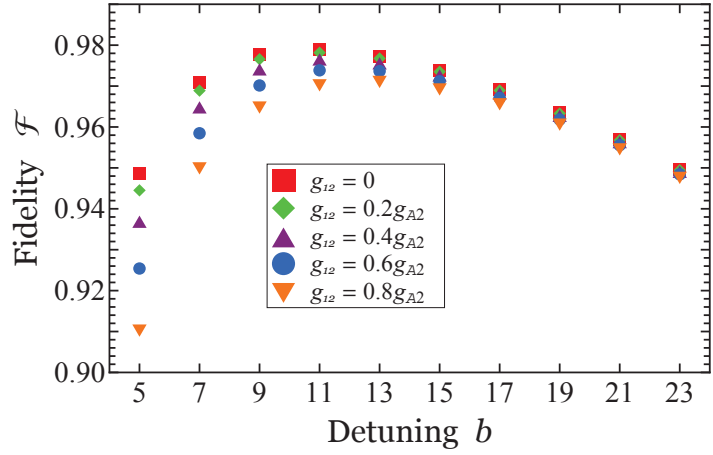

FIG. 3: (Color online) Fidelity of the entanglement preparation versus the normalized detuning $b = |\delta_1|/g_1$. Refer to the text for the parameters used in the numerical calculation.

mode of cavity $j$ and the $|1\rangle \leftrightarrow |2\rangle$ transition of qutrit $j$, with coupling constant $\widetilde{g}_j$ and detuning $\widetilde{\delta}_j = \omega_{21j} - \omega_{c_j}$ [Fig. 2(a)], while the second term indicates the unwanted off-resonant coupling between the mode of cavity $j$ and the $|1\rangle \leftrightarrow |2\rangle$ transition of qutrit $A$, with coupling constant $\widetilde{g}_{Aj}$ and detuning $\widetilde{\delta}_{Aj} = \omega_{21A} - \omega_{c_j}$ [Fig. 2(b)]. It should be mentioned that the term describing the cavity-induced coherent $|0\rangle \leftrightarrow |2\rangle$ transition for each qutrit is not included in the Hamiltonians $H'_I$, since this transition is negligible because of $\omega_{cj} \ll \omega_{20j}, \omega_{20A}$ ($j = 1, 2$) (Fig. 2). The last term describes the intercavity crosstalk betwee the two cavities, with $\Delta = \omega_{c2} - \omega_{c1} = \delta_{A1} - \delta_{A2}$.

The dynamics of the lossy system, with finite qutrit relaxation and dephasing and photon lifetime included, is determined by

$$\begin{aligned}
\frac{d\rho}{dt} &= -i\left[\widetilde{H}_I, \rho\right] + \sum_{j=1}^{2} \kappa_j \mathcal{L}\left[\hat{a}_j\right] \\
&+ \sum_{j=1,2,A} \left\{\gamma_j \mathcal{L}\left[\sigma_j^-\right] + \gamma_{21j}\mathcal{L}\left[\sigma_{21j}^-\right] + \gamma_{20j}\mathcal{L}\left[\sigma_{20j}^-\right]\right\} \\
&+ \sum_{j=1,2,A} \left\{\gamma_{j,\varphi 1}\left(\sigma_{11j}\rho\sigma_{11j} - \sigma_{11j}\rho/2 - \rho\sigma_{11j}/2\right)\right\} \\
&+ \sum_{j=1,2,A} \left\{\gamma_{j,\varphi 2}\left(\sigma_{22j}\rho\sigma_{22j} - \sigma_{22j}\rho/2 - \rho\sigma_{22j}/2\right)\right\},
\end{aligned} \qquad (18)$$

where $\sigma_{20j}^- = |0\rangle_j \langle 2|, \sigma_{20A}^- = |0\rangle_A \langle 2|, \sigma_{11j} = |1\rangle_j \langle 1|, \sigma_{22j} = |2\rangle_j \langle 2|$; and $\mathcal{L}[\Lambda] = \Lambda\rho\Lambda^+ - \Lambda^+\Lambda\rho/2 - \rho\Lambda^+\Lambda/2$, with $\Lambda = \hat{a}_j, \sigma_j^-, \sigma_{21j}^-, \sigma_{20j}^-$. In addition, $\kappa_j$ is the photon decay rate of cavity $a_j$, $\gamma_j$ is the energy relaxation rate of the level $|1\rangle$ of qutrit $j$, $\gamma_{21j}$ ($\gamma_{20j}$) is the energy relaxation rate of the level $|2\rangle$ of qutrit $j$ for the decay path $|2\rangle \to |1\rangle$ ($|0\rangle$), and $\gamma_{j,\varphi 1}$ ($\gamma_{j,\varphi 2}$) is the dephasing rate of the level $|1\rangle$ ($|2\rangle$) of qutrit $j$.

The fidelity of the operation is given by

$$\mathcal{F} = \langle\psi_{\text{id}}|\widetilde{\rho}|\psi_{\text{id}}\rangle, \qquad (19)$$

where $|\psi_{\text{id}}\rangle$ is the output state of an ideal system (i.e., without dissipation, dephasing, and crosstalks) as discussed in the previous section; and $\widetilde{\rho}$ is the final density operator of the system when the operation is performed in a realistic physical system.

For entanglement preparation, $|\psi_{\text{id}}\rangle$ is $|\phi\rangle|0\rangle_{c1}|0\rangle_{c2}$; while for information transfer, it is the state $|\varphi\rangle|0\rangle_{c1}|0\rangle_{c2}$. Here and above, $|0\rangle_{cj}$ is the vacuum state of cavity $j$ ($j = 1, 2$).

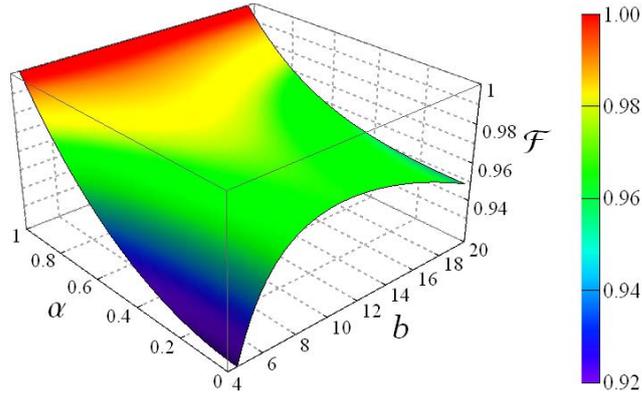

FIG. 4: (Color online) Fidelity of the information transfer versus $(b, \alpha)$. Here, the detuning is $b = |\delta_1|/g_1$, and $\alpha = \sqrt{1-\beta^2}$. For simplicity, here we consider the transferred state $\alpha |0\rangle + \beta |1\rangle$, with real numbers $\alpha$ and $\beta$. For the parameters used in the numerical calculation, see the text.

### A. Fidelity for the entanglement preparation

Without loss of generality, let us consider three identical superconducting phase qutrits. According to the condition (4), we set $\delta_1/(2\pi) = \delta_{A1}/(2\pi) = -0.5$ GHz and $\delta_2/(2\pi) = \delta_{A2}/(2\pi) = -1$ GHz. For the setting here, we have $\Delta/2\pi = 0.5$ GHz. Set $\widetilde{\delta}_j = \delta_j - 0.05\omega_{10j}$ and $\widetilde{\delta}_{Aj} = \delta_{Aj} - 0.05\omega_{10A}$ ($j = 1, 2$) [49]. For superconducting phase qubits, the typical qubit transition frequency is between 4 and 10 GHz. Thus, we choose $\omega_{10A}/2\pi, \omega_{10j}/2\pi \sim 6.5$ GHz. Note that $g_2$ is determined based on Eq. (10), given $\delta_1, \delta_2$, and $g_1$; and thus the ratio of $\delta_2/g_2$ can be calculated, if $\delta_2$ and $g_2$ are known. In addition, $g_{A1}$ and $g_{A2}$ are determined based on Eq. (10), given $\delta_1$ and $\delta_2$. Next, one has $\widetilde{g}_j \sim \sqrt{2}g_j$ and $\widetilde{g}_{Aj} \sim \sqrt{2}g_{Aj}$ ($j = 1, 2$) for the phase qubit here. For example, we choose $\gamma_{j,\varphi 1}^{-1} = \gamma_{j,\varphi 2}^{-1} = 2.5$ $\mu$s, $\gamma_j^{-1} = 10$ $\mu$s, $\gamma_{21j}^{-1} = 7.5$ $\mu$s, and $\gamma_{20j}^{-1} = 30$ $\mu$s; and $\kappa_1^{-1} = \kappa_2^{-1} = 5$ $\mu$s. For a phase qutrit with the three levels considered here, the $|0\rangle \leftrightarrow |2\rangle$ dipole matrix element is much smaller than that of the $|0\rangle \leftrightarrow |1\rangle$ and $|1\rangle \leftrightarrow |2\rangle$ transitions. Thus, $\gamma_{20j}^{-1} \gg \gamma_{10j}^{-1}, \gamma_{21j}^{-1}$.

For the parameters chosen above, the fidelity versus $b = |\delta_1|/g_1$ is plotted in Fig. 3 for $g_{12} = 0, 0.2g_{\max}$, $0.4g_{\max}, 0.6g_{\max}, 0.8g_{\max}$. From Fig. 3, one can see that for $g_{12} \leq 0.2g_{\max}$, the effect of intercavity cross coupling between the two cavities on the fidelity of the operation is negligible, which can be seen by comparing the top two curves. Moreover, Fig. 3 shows that for $b \sim 11$ and $g_{12} = 0.2g_{\max}$, a high fidelity $\sim 98\%$ is available for the entanglement preparation.

### B. Fidelity for the information transfer

The parameters used in the numerical calculation are the same as above. Fig. 4 shows the fidelity versus $(b, \alpha)$, which is plotted for $g_{12} = 0.2g_{\max}$. One can see from Fig. 4 that for $b \sim 9$, a high fidelity $> 97\%$ is achievable for the information transfer. Further, it is predicted that a higher fidelity can be obtained when $g_{12} < 0.2g_{\max}$.

This condition, $g_{12} \leq 0.2g_{\max}$, is not difficult to satisfy with the typical capacitive cavity-qutrit coupling illustrated in Fig. 1(a). As long as the cavities are physically well separated, the intercavity cross-talk coupling strength is $g_{12} \sim g_{A1}C_2/C_\Sigma, g_{A2}C_1/C_\Sigma$, where $C_\Sigma = C_1 + C_2 + C_q$. For $C_1, C_2 \sim 1$ fF and $C_\Sigma \sim 10^2$ fF (the typical values of the cavity-qutrit coupling capacitance and the sum of all coupling capacitance and qutrit self-capacitance, respectively), we have $g_{12} \sim 0.01g_{A1}, 0.01g_{A2}$. Note that $g_{Aj} \leq g_{\max}$. Thus, the condition $g_{12} \leq 0.2g_{\max}$ can be readily met in experiments. Hence, implementing designs with sufficiently weak direct intercavity couplings is straightforward.

For $b \sim 11$, we have $\{g_1, g_2, g_{A1}, g_{A2}\} \sim \{45.5, 64.3, 32.2, 45.5\}$ MHz. Note that a coupling constant $\sim 220$ MHz can be reached for a superconducting qutrit coupled to a one-dimensional CPW (coplanar waveguide) resonator [35], and that $T_1$ and $T_2$ can be made to be on the order of $10 - 100$ $\mu$s or longer for state-of-the-art superconducting devices [50]. The energy relaxation time $T_1'$ and dephasing time $T_2'$ of the level $|2\rangle$ are comparable to $T_1$ and $T_2$, respectively. For instance, $T_1' \sim T_1/\sqrt{2}$ and $T_2' \sim T_2$ for phase qutrits. For $\omega_{10A}/2\pi, \omega_{10j}/2\pi \sim 6.5$ GHz chosen above, we have $\omega_{c1}/2\pi \sim 6$ GHz and $\omega_{c2}/2\pi \sim 5.5$ GHz. For the cavity frequencies chosen here and the values of $\kappa_1^{-1}$ and $\kappa_2^{-1}$ used in the numerical calculation, the required quality factors for the two cavities are $Q_1 \sim 1.9 \times 10^5$ and $Q_2 \sim 1.7 \times 10^5$, respectively. Note that superconducting CPW resonators with a loaded quality factor $Q \sim 10^6$ have been experimentally demonstrated [51,52], and planar superconducting resonators with internal quality factors above one million ($Q > 10^6$) have also been recently reported [53]. Our analysis given here demonstrates that high-fidelity implementation of the entangled state and the information transfer by using this proposal is feasible within the present circuit QED technique. We remark that further investigation is needed for each particular experimental setup. However, this requires a rather lengthy and complex analysis, which is beyond the scope of this theoretical work.

## IV. CONCLUSION

We have proposed a method to generate quantum entanglement and perform quantum information transfer between two spatially-separate superconducting qubits residing in two different cavities. As shown above, this work is of interest because the entanglement generation and information transfer implementation do not require employing photons of the cavities as quantum buses and thus decoherence caused due to the cavity decay is greatly supressed during the entire operation. The proposal does not require applying classical microwave pulses and needs only one step of operation and one superconducting coupler qubit, so that the circuit complexity is much reduced and the operation is greatly simplified. In addition, our analysis shows that high-fidelity implementation of this proposal with superconducting phase qubits is feasible within the present circuit QED technology. Finally, it is noted that the method presented here is quite general, and can be applied to accomplish the same task with the coupler qubit replaced by a different type of qubit such as a quantum dot, or with the two intercavity qubits replaced by other two qubits, e.g., two atoms, two quantum dots, two NV centers, and so on.

## ACKNOWLEDGMENTS

We thank Shi-Biao Zheng for many fruitful discussions. C.P.Y. was supported in part by the National Natural Science Foundation of China under Grant No. 11074062, the Zhejiang Natural Science Foundation under Grant No. LZ13A040002, and the funds from Hangzhou Normal University under Grant No. HSQK0081. Q.P.S. was supported by the National Natural Science Foundation of China under Grant No. 11147186. This work is also partially supported by the ARO, RIKEN iTHES Project, MURI Center for Dynamic Magneto-Optics, JSPS-RFBR contract No. 12-02-92100, Grant-in-Aid for Scientific Research (S), MEXT Kakenhi on Quantum Cybernetics, and the JSPS via its FIRST program.

[1] Cirac J I and Zoller P 1995 *Phys. Rev. Lett.* **74** 4091
[2] Blatt R and Wineland D 2008 *Nature* **453** 1008
[3] Duan L M, Lukin M D, Cirac J I and Zoller P 2001 *Nature* **414** 413
[4] Chou C W, Laurat J, Deng H, Choi K S, Riedmatten H D, Felinto D, and Kimble H J 2007 *Science* **316** 1316
[5] Mabuchi H and Doherty A C 2002 *Science* **298** 1372
[6] Hagley E, Maître X, Nogues G, Wunderlich C, Brune M, Raimond J M, and Haroche S 1997 *Phys. Rev. Lett.* **79** 1
[7] Zheng S B and Guo G C 2000 *Phys. Rev. Lett.* **85** 2392
[8] Osnaghi S, Bertet P, Auffeves A, Maioli P, Brune M, Raimond J M, and Haroche S *et al.* 2001 *Phys. Rev. Lett.* **87** 037902
[9] Cirac J I, Zoller P, Kimble H J, and Mabuchi H 1997 *Phys. Rev. Lett.* **78** 3221; Pellizzari T 1997 *Phys. Rev. Lett.* **79** 5242
[10] Ye S Y, Zhong Z R, and Zheng S B 2008 *Phys. Rev. A* **77** 014303; Serafini A, Mancini S, and Bose S 2006 *Phys. Rev. Lett.* **96** 010503; Yin Z Q and Li F L 2007 *Phys. Rev. A* **75** 012324; Lü X Y, Liu J B, Ding C L, and Li J H 2008 *Phys. Rev. A* **78** 032305; Lü X Y, Song P J, Liu J B, and Yang X 2009 *Opt. Express* **17** 14298; Dong Y L, Zhu S Q, and You W L 2012 *Phys. Rev. A* bf 85 023833
[11] Ritter S, Nölleke C, Hahn C, Reiserer A, Neuzner A, Uphoff M, Mücke M, Figueroa E, Bochmann J, and Rempe G 2012 *Nature* **484** 195
[12] You J Q and Nori F 2005 *Physics Today* **58** (11) 42; You J Q and Nori F 2011 *Nature* **474** 589
[13] Xiang Z L, Ashhab S, You J Q, and Nori F 2013 *Rev. Mod. Phys.* **85** 623
[14] Buluta I, Ashhab S, and Nori F 2011 *Reports on Progress in Physics* **74** 104401; Shevchenko S N, Ashhab S, and Nori F 2010 *Phys. Reports* **492** 1; Nation P D, Johansson J R, Blencowe M P, and Nori F 2012 *Rev. Mod. Phys.* **84** 1
[15] Wallraff A, Schuster D I, Blais A, Frunzio L, Huang R S, Majer J, Kumar S, Girvin S M, and Schoelkopf R J 2004 *Nature* **431** 162
[16] Houck A A *et al.* 2007 *Nature* **449** 328
[17] Marquardt F and Bruder C 2001 *Phys. Rev. B* **63** 054514

[18] Liu Y X, Wei L F, and Nori F 2004 *Europhys. Lett.* **67** 941
[19] Marquardt F 2007 *Phys. Rev. B* **76** 205416
[20] Mariantoni M, Storcz M J, Wilhelm F K, Oliver W D, Emmert A, Marx A, Gross R, Christ H, and Solano E arXiv:cond-mat/0509737.
[21] Hofheinz M, Weig E M, Ansmann M, Bialczak R C, Lucero E, Neeley M, OConnell A D, Wang H, Martinis J M, and Cleland A N 2008 *Nature* **454** 310; Wang H, Hofheinz M, Ansmann M, Bialczak R C, Lucero E, Neeley M, OConnell A D, Sank D, Wenner J, Cleland A N, and Martinis J M 2008 *Phys. Rev. Lett.* **101** 240401
[22] Hofheinz M, Wang H, Ansmann M, Bialczak R C, Lucero E, Neeley M, O'Connell A D, Sank D, Wenner J, Martinis J M, and Cleland A N 2009 *Nature* **459** 546
[23] Yang C P, Chu S I, and Han S 2003 *Phys. Rev. A* **67** 042311
[24] You J Q and Nori F 2003 *Phys. Rev. B* **68** 064509
[25] Blais A, Huang R S, Wallraff A, Girvin S M, and Schoelkopf R J 2004 *Phys. Rev. A* **69** 062320
[26] Yang C P, Chu S I, and Han S 2004 *Phys. Rev. Lett.* **92** 117902
[27] Plastina F and Falci G 2003 *Phys. Rev. B* **67** 224514
[28] Blais A, Maassen van den Brink A, and Zagoskin A M 2003 *Phys. Rev. Lett.* **90** 127901
[29] Helmer F and Marquardt F 2009 *Phys. Rev. A* **79** 052328
[30] Bishop L S *et al.* 2009 *New J. Phys.* **11** 073040
[31] Yang C P, Liu Y X, and Nori F 2010 *Phys. Rev. A* **81** 062323
[32] Yang C P, Zheng S B, and Nori F 2010 *Phys. Rev. A* **82** 062326
[33] Majer J *et al.* 2007 *Nature* **449** 443
[34] Leek P J, Filipp S, Maurer P, Baur M, Bianchetti R, Fink J M, Goppl M, Steffen L, and Wallraff A 2009 *Phys. Rev. B* **79** 180511(R)
[35] DiCarlo L *et al.* 2010 *Nature* **467** 574
[36] Mariantoni M *et al.* 2011 *Science* **334** 61
[37] Fedorov A, Steffen L, Baur M, Silva M P, and Wallraff A 2012 *Nature* **481** 170
[38] Mariantoni M, Deppe F, Marx A, Gross R, Wilhelm F K, and Solano E 2008 *Phys. Rev. B* **78** 104508
[39] Strauch F W, Jacobs K, and Simmonds R W 2010 *Phys. Rev. Lett.* **105** 050501
[40] Merkel S T and Wilhelm F K 2010 *New J. Phys.* **12** 093036
[41] Wang H *et al.* 2011 *Phys. Rev. Lett.* **106** 060401
[42] Yang C P, Su Q P, and Han S 2012 *Phys. Rev. A* **86** 022329
[43] Yang C P, Su Q P, Zheng S B, and Han S 2013 *Phys. Rev. A* **87** 022320
[44] Zheng Z F, Su Q P, and Yang C P 2013 *J. Phys. Soc. Jpn.* **82** 084801
[45] Zheng S B 2011 *Phys. Rev. Lett.* **87** 230404
[46] Clarke J and Wilhelm F K 2008 *Nature* **453** 1031
[47] Neeley M, Ansmann M, Bialczak R C, Hofheinz M, Katz N, Lucero E, OConnell A, Wang H, Cleland A N, and Martinis J M 2008 *Nature Phys.* **4** 523; Zagoskin A M, Ashhab S, Johansson J R, and Nori F 2006 *Phys. Rev. Lett.* **97** 077001
[48] Sandberg M, Wilson C M, Persson F, Bauch T, Johansson G, Shumeiko V, Duty T, and Delsing P 2008 *Appl. Phys. Lett.* **92** 203501
[49] For a phase qutrit, a ratio 5% of the anharmonicity between the $|0\rangle \leftrightarrow |1\rangle$ transition frequency and the $|1\rangle \leftrightarrow |2\rangle$ transition frequency to the the $|1\rangle \leftrightarrow |2\rangle$ transition frequency is readily achieved in experiments.
[50] see, Bylander J *et al.* 2011 *Nature Phys.* **7** 565; Paik H 2011 *Phys. Rev. Lett.* **107** 240501; Chow J M *et al.* 2012 *Phys. Rev. Lett.* **109** 060501; Rigetti C *et al.* 2012 *Phys. Rev. B* **86** 100506(R); R. Barends *et al.*, arXiv:1304.2322
[51] Chen W, Bennett D A, Patel V, and Lukens J E 2008 *Supercond. Sci. Technol.* **21** 075013
[52] Leek P J, Baur M, Fink J M, Bianchetti R, Steffen L, Filipp S, and Wallraff A 2010 *Phys. Rev. Lett.* **104** 100504
[53] Megrant A *et al.* 2012 *Appl. Phys. Lett.* **100** 113510